# Real-time monitoring of single-photon detectors against eavesdropping in quantum key distribution systems

Thiago Ferreira da Silva, [1,2,*] Guilherme B. Xavier, [3,4,5]
Guilherme P. Temporão, [1] and Jean Pierre von der Weid[1]

[1] *Centre for Telecommunication Studies, Pontifical Catholic University of Rio de Janeiro, R. Marquês de São Vicente 225 Gávea, 22451-900 Rio de Janeiro, Brazil*
[2] *Optical Metrology Division, National Institute of Metrology, Quality and Technology, Av. Nossa Sra. das Graças 50, 25250-020 Duque de Caxias, RJ, Brazil*
[3] *Departamento de Ingeniería Eléctrica, Universidad de Concepción, Casilla 160-C, Correo 3, Concepción, Chile*
[4] *Center for Optics and Photonics, Universidad de Concepción, Casilla 4016, Concepción, Chile*
[5] *MSI-Nucleus on Advanced Optics, Universidad de Concepción, Casilla 160-C, Concepción, Chile*
[*] *thiago@opto.cetuc.puc-rio.br*

**Abstract:** By employing real-time monitoring of single-photon avalanche photodiodes we demonstrate how two types of practical eavesdropping strategies, the after-gate and time-shift attacks, may be detected. Both attacks are identified with the detectors operating without any special modifications, making this proposal well suited for real-world applications. The monitoring system is based on accumulating statistics of the times between consecutive detection events, and extracting the afterpulse and overall efficiency of the detectors in real-time using mathematical models fit to the measured data. We are able to directly observe changes in the afterpulse probabilities generated from the after-gate and faint after-gate attacks, as well as different timing signatures in the time-shift attack. We also discuss the applicability of our scheme to other general blinding attacks.

## 1. Introduction

Quantum Key Distribution (QKD) [1] has gained widespread fame for its ability to reliably share a secret random string of bits between two remotely separated parties, typically known as Alice and Bob. This string can then be used as a key to implement the one-time pad, a cryptographic scheme able to provide unconditional security against an eavesdropper, Eve. The security of QKD has been theoretically proven against a wide class of attacks, even when taking into account the imperfection of the devices [1-4]. The issue, however, rests in the fact that real implementations with current technology presents loopholes that can be exploited by Eve, without her presence being directly revealed to Alice and Bob.

There are many different device imperfections that Eve may base her attacks on, such as a nonzero multi-photon emission probability from the optical source (through a photon-number splitting attack [5]), or with an imperfect Faraday mirror in the plug & play system [6], thus compromising security. Alternatively, Eve can even act directly probing on the system components, like in the Trojan horse attack [7], or she can fake the phase encoding of a bidirectional system, as in the phase-remapping strategy [8,9]. However, most of the recently proposed "quantum hacking" schemes focus on the single-photon detectors, which have been shown to perform quite differently from the ideal behaviour.

For example, by acting on the system timing, Eve can implement the class of time-shift attacks [10-12]. The attack is based on the non-uniformity of the detection efficiency throughout the detection gate in real detectors. By shifting the time of arrival of the single photons inside the detector gate window, Eve can implement her attack and infer the results on Bob's measurements. The attack may include an intercept-resend strategy sending faked states to Bob [10] or just a random delay applied to the timing of the qubits [11,12]. In addition, depending on the synchronization protocol performed by Alice and Bob before starting the key distribution, the detector efficiency mismatch can be artificially enhanced by Eve [13].

Another class of attacks is based upon the direct control of the detectors at Bob's station, typically employing classical light to somehow manipulate their response, by forcing/suppressing an avalanche according to Eve's will. The detectors can be temporarily blinded to single photons due to an optically induced drop in the diode bias voltage [14-16] or with a thermally induced rise in the breakdown voltage [17]. Eve intercepts the qubits and sends faked states [18] to Bob as bright pulses, allowing her to obtain total control of the detectors without (in principle) being caught. The deadtime of the detectors can also be used by Eve to eavesdrop on the cryptographic key without intercepting the photons sent by Alice [19], through optically blinding the detector due to the induced deadtime.

Attacks aimed on imperfections of the single-photon detector have been demonstrated against commercial systems [20], in a full-field implementation [21] and have even showed that it is possible to fake the violation of Bell´s inequalities [22]. All these recent attacks are a real issue because they can all be done with current technology. There has been some discussion regarding whether using a low impedance bias voltage source or appropriately setting the detector discrimination levels would avoid blinding attacks [23-26]. As alternative countermeasures, the usage of optical power meters at the entrance of Bob's apparatus or even monitoring of the APD current and temperature have been proposed [24, 27]. Recently a new idea was proposed in order to render the system immune to all detector side-channel attacks, called measurement-device independent quantum-key distribution (MDI-QKD) [28]. It requires a Bell-state measurement of two independent remote faint laser sources, thus being experimentally challenging and requires a different topological setup for QKD systems.

In this paper we propose a different approach based on real-time characterization of single-photon detectors [29], which may be readily applied to current QKD setups. We are able to observe the fingerprints left on the detectors nominal characteristics by the after-gate [16] and time-shift [11] attacks, thus revealing Eve's presence. This method does not require any hardware modifications such as optical taps, extra power meters or single-photon detectors [17], nor access to the APD parameters as temperature and current [14,24]. In fact it can be done only through software real-time processing of the single-photon avalanche photodiodes (SPAD) output signal as long as an accurate timing unit is present in Bob's detection apparatus, also making it a very attractive and cost-effective countermeasure for eavesdropping in QKD.

In the next section we briefly review how our characterization method works, then in section 3 we experimentally implement the after-gate attack and demonstrate the feasibility of catching Eve using real-time characterization for several different operating conditions. We also test our method against the recently proposed "faint after-gate" [30] attack, a more subtle invasive technique, and find that it is effective under certain conditions. Then we discuss the applicability of the monitoring system against more general blinding attacks. In section 4 we monitor a single-photon detector subjected to a proof-of-principle time-shift attack and in section 5 the conclusions are presented.

## 2. Real-time SPAD characterization

The SPAD [31,32] is a key element in all commercial quantum communication schemes, therefore their correct characterization is of great interest for reliable operation of QKD protocols. Three important characterization parameters are the overall detection efficiency ($\eta$), the dark count ($P_d$) and afterpulse probabilities ($P_{after}$). All these parameters can be characterized before the detector is put into operation, but may vary under different operational conditions. We have shown that it is possible to fully simultaneously characterize these three parameters in real-time, with the detector under normal operation as part of a quantum communication system [29]. This is in contrast to previous characterization methods, in which the detector had to be under specific conditions [33,34], and therefore unable to be performed under a real operating situation.

An analytical mathematical model has been derived describing the probabilities that a count event registered by the SPAD has been caused by a single photon arriving on the detector, a dark count or by an afterpulse, as a result of a previous avalanche. This model is expressed as a function of the time interval $m$ between consecutive detections of the SPAD, regardless of whether it is a "true" count generated by the absorption of a single-photon ($1-P_{np}$), or a noise count coming from a dark count ($P_{dark}$) or afterpulse ($P_a(m)$) [29]:

$$P(m) = \{1 - P_{np} \cdot (1 - P_{dark}) \cdot [1 - P_a(m)]\} \times [P_{np} \cdot (1 - P_{dark})]^{m-1} \times [1 - \alpha(m) + \beta(m)], \quad (1)$$

where $P_{np} = exp(-\mu\eta)$ is the probability of having zero photons in a detection gate assuming a Poissonian photon number distribution, with $\mu$ representing the average number of photons per pulse (or detector gate window) and $\eta$, the overall detection efficiency. The terms $\alpha(m)$ and $\beta(m)$ are expressed as functions of exponentials of multiples of the time interval [29].

The method, as depicted in Fig. 1, works as follows: the times between consecutive detections are accumulated by a computer using an analog-to-digital converter (A/D) acquisition board and a histogram is plot, which is equivalent to the probability distribution of the detection events. Eq. (1) is fit to the normalized histogram in real-time and the parameters $P_d$, $P_0$, $\tau$ and the product $\mu\eta$ are simultaneously obtained. The method can also determine the detector deadtime (at the A/D resolution) directly from the histogram.

The total afterpulse probability $P_{after}$ is obtained from the summation of $P_a$ from $m=1$ to infinity, and is calculated from the measured parameters as:

$$P_{after} = P_0 \frac{1}{exp(T/\tau) - 1}. \quad (2)$$

The error in the obtained parameters clearly depends on the measurement sample length. We found that the $R^2$ parameter of the fit was better than 0.99 for $3\times10^4$ detections (the closer to unity the more accurate the fit is), and the standard deviation, relative to the absolute measured value, is 0.3% for the overall detection efficiency and 5.0% for the total afterpulse probability for the same number of samples [29]. Therefore Bob is able to obtain reliable information on the detector's parameters for a small number of detections. If he finds his detector is operating outside its nominal values, he can discard those bits accumulated and report to Alice that an eavesdropper may be attempting to control his detectors. We shall see in the next sections that the attacks leave marks on the detectors in the form of abnormal behavior, which is caught with our real-time system. For further details on this system we advise the reader to look up [29].

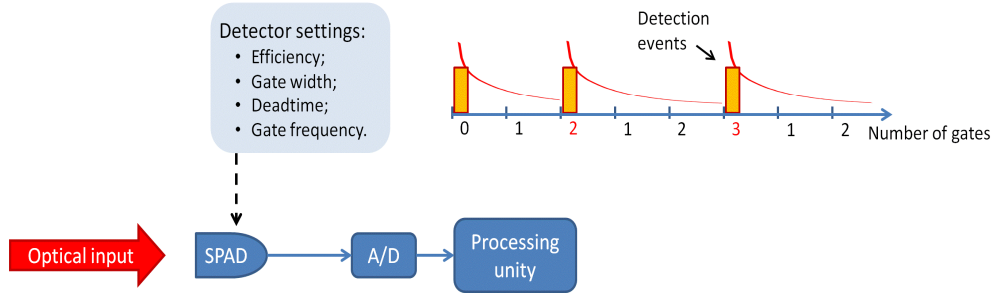

Fig. 1. Real-time SPAD monitoring system. The decaying exponentials qualitatively illustrate the afterpulse probability in subsequent gate windows.

## 3. Monitoring the SPAD against blinding attacks

*3.1 After-gate attack*

In the so-called "after-gate attack" [16], the eavesdropper externally takes control over Bob's gated Geiger-mode [1,35] SPADs. Eve sends bright optical pulses to Bob, prepared in the desired basis according to her results on the measurement of the intercepted qubits. The pulse forces an avalanche with 100% probability whenever Bob's basis choice matches Eve's. This behaviour is obtained with a correct tuning of the bright pulse intensity and explores the linear regime of the SPAD when outside of the detection time window. If Eve and Bob's bases match, the optical pulse reaching the SPAD has sufficient power to exceed the avalanche breakdown causing a photon count. Otherwise, the pulse's optical power is split and is not sufficient to trigger an avalanche. Using this strategy, Eve remains (at least in principle) undetected, as no error is introduced.

The bright pulses are sent after a detection gate to minimize afterpulse probability, because even if a bright pulse reaching the detector outside of a gate is not able to trigger an avalanche, traps are filled due to the linear operation of the device. Provided a long deadtime or low gate rate are not imposed, the probability of an afterpulse to occur in the next gates becomes higher, as the main detrapping lifetime of an InGaAs-based gated SPAD is typically of few μs [29,33].

In a complete BB84 system, Alice individually prepares the single photons in one out of four states, randomly chosen from a pair of non-orthogonal bases [1]. Bob then randomly chooses one of the two bases to measure each qubit. In the case of polarisation encoding, the basis choice and measurement can be implemented with an active element that performs polarisation rotations followed by a polarising beamsplitter (PBS), which has one SPAD connected to each output port. According to the preparation and measurement sets, one of the SPADs will click.

The eavesdropper detection apparatus is similar to Bob's, having a basis choice element, a PBS and a pair of SPADs. Eve intercepts the photons sent by Alice, measures them in a random basis, and sends a bright optical pulse to Bob, with polarisation encoded in one out of the four BB84 states, according to the measurement result. After Bob's basis choice, the pulse will cause the correspondent detector to click (if the bases agree) or the pulse may be split and reach both detectors with half the power (if the bases disagree), not causing an avalanche in neither of them. Even by sending the pulses after the detection gate, the main drawback of this attack is that it will nevertheless increase the afterpulse probability of the detectors [16], and this abnormal behaviour is what we look for in our measurement.

For the experimental demonstration a simplified version of a BB84-based QKD system with polarisation encoding was built, together with the eavesdropper apparatus, according to Fig. 2. A commercial SPAD was monitored with our method while submitted to the after-gate attack under different conditions.

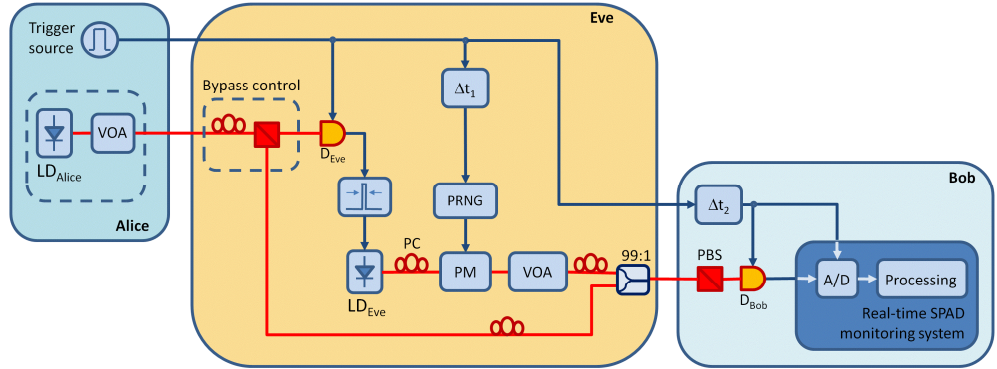

Fig. 2. Experimental setup for the after-gate attack. Red lines represent optical fibre connections. The trigger source is placed under Alice's setup indicating that she is providing the synchronization signals for all the parties involved. A/D: Analog-to-digital converter; D: Single-photon detector; LD: Laser diode; PBS: Polarising beam-splitter; PC: Polarisation controller; PM: Polarisation modulator; PRNG: Pseudo-random number generator; VOA: Variable optical attenuator; Δt: delay generator.

In our setup, Alice's station is composed of a single-photon source made of a CW (continuous-wave) telecom laser diode ($LD_{Alice}$) with a variable optical attenuator (VOA). She also has an electrical trigger source, which is used to synchronize the entire QKD system, as in a real setup. The photons sent by Alice are intercepted by Eve, which has one commercial SPAD module ($\eta$=15% and 2.5 ns gate width) with detection gates driven by Alice's trigger pulses. In a realistic QKD setup Alice's laser would be pulsed, a fact that does not affect the results because every time Eve's detector is opened (as dictated by Alice's trigger) there is an average of $\mu$ photons on that detection window, with probability of detection $\eta$. Through the delay control to Bob, Eve is able to make sure her bright pulses or the bypassed photons (through the bypass control, see below for more details) arrive at Bob with the proper timing. Each detection event at Eve creates a 102.5 ns wide electrical pulse that, after compression to around 1 ns width, directly modulates her laser diode ($LD_{Eve}$) to create a bright optical pulse. An additional simplification is that Eve is simply detecting the single-photons without any polarisation analysis, since Alice is only sending a single state of polarisation from BB84, a fact that does not affect the measurement results. Eve, therefore, prepares the attack pulse's polarisation state according to a pseudo-random number generator (PRNG), simulating her basis choice that would be dependent on her random measurements on the states sent by Alice. Eve chooses between two maximally non-orthogonal polarisation states, corresponding to states on two different encoding bases. The binary random sequence, read from FPGA-based electronics, is applied as a bi-stable voltage to drive a $LiNbO_3$ polarization modulator

(PM) [36]. We are also assuming that both bases are equally likely to be used by Alice and Bob. The polarisation changes are synchronized with the optical pulses through a delayed version of the main trigger signal ($\Delta t_1$). The intensity of the bright pulses is adjusted with a variable optical attenuator (VOA) according to the avalanche threshold of Bob's detector outside the detection gate, previously characterized at a low repetition rate by scanning the optical pulse relative to the temporal gate window at different optical powers.

At the receiving station, a PBS is inserted in front of Bob's detector ($D_{Bob}$). We are using a single detector to demonstrate the attack and the application of the real-time monitoring system against eavesdropping. Depending on Eve's basis choice, hers and Bob's bases may agree, and the bright pulse causes the detector to click; otherwise, it causes no detection event. A delayed version of the synchronization signal ($\Delta t_2$) is used to trigger the SPAD under attack. Eve sends the bright pulses such that they arrive just after the end of Bob's detector gate windows, adjusted with the delay line $\Delta t_2$. The A/D board of our monitoring system is connected to the electrical output of Bob's detector and to the trigger signal. The statistics of times between consecutive detection events are acquired, using the number of opened gates as the timebase, and analysed in real-time.

Every time Eve's and Bob's bases disagree, the current flow generated by the absorbed optical pulse through Bob's detector resets the time-dependent afterpulse exponential probability, but without immediately triggering an avalanche. This process causes the mathematical model that generated Eq. (1) to become intractable (several afterpulse terms with different probabilities of occurring appear in Eq. (1) depending on Alice and Bob´s choices) and we use a different procedure to extract the afterpulse information from the data collected. We obtain the total afterpulse probability from the measured data, by fitting a straight line to the long tail of the logarithm of the normalized histogram of time counts. The final value, calculated in real-time, is taken from the area above this line as compared to the total histogram area normalized to unity [28]. Note that this method is only used to extract the afterpulse probability. The other parameters (dependent on the slope and offset of the distribution) may still be directly extracted from the measured data using Eq. (1) without the afterpulse contributions.

We have also simulated the fact that Eve may not intercept all the single-photons sent from Alice to Bob, in order to try to mask her presence at the cost of a reduced portion of acquired information. This can be adjusted by Eve at the bypass control shown in Fig. 2. By rotating the polarisation controller (PC) in front of the PBS, Eve can bypass a fraction of the total number of photons to Bob and attack the complementary number. Eve recombines her optical pulses and the single-photons from Bob with a 99:1 splitter such that the lower transmittance arm is used to pass through the bright pulses, whose power is adjusted to overcome this loss. Alice's laser is attenuated to have 0.4 photons on average per 2.5 ns window, which is also the effective gate width used by Bob. His detector was previously characterized at 400 kHz of gate frequency with 10 µs of dead time and 15% detection efficiency, in order to compare its characterization signature with and without the attack. This dead time was applied since it is typical for Bob to employ one on his detectors in QKD systems to minimize the error rate generated from afterpulses.

The histograms of times for Bob's detector were obtained at different ratios of eavesdropped per transmitted photons and are depicted in Fig. 3, with time represented by number of gates between consecutive detections. The average photon number dependence is reflected mainly by the slope of the long curve tail [28]. From the figure we see that the average number of photons per pulse detected by Bob does not change (parallel curves). This is important as an attacking strategy for Eve because, on the contrary, an extra fingerprint would be left by her. If the average photon number increases, there is a higher probability that at least one photon is encountered in a gated time window, as can be verified in the Poissonian distribution of a faint laser. This makes the events closer to each other and the slope of the curve increases.

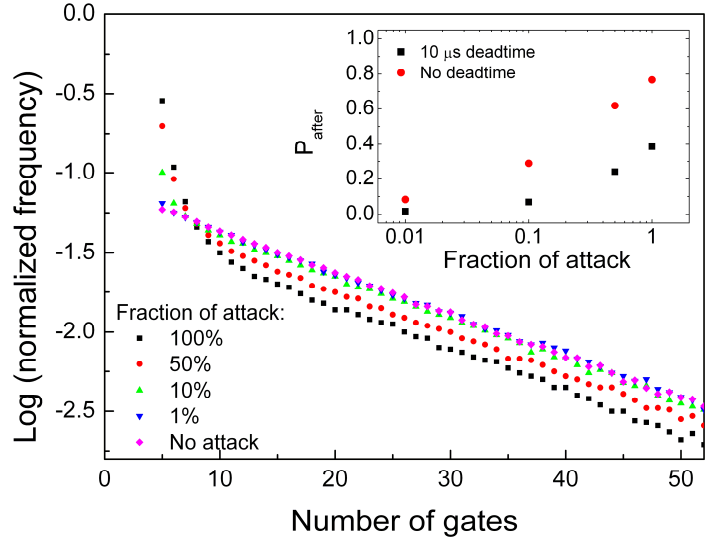

Fig. 3. Comparative histograms with SPAD behaviour under different fractions of eavesdropped photons. Inset highlights the afterpulse contribution, which occurs largely at short time intervals, as a function of the fraction of eavesdropped photons. No counts are generated in the first bins due to the imposed deadtime of 10 μs. In all cases the eavesdropper intervention can be detected with our monitoring system.

The sharp region in the beginning of the histogram, highlighted in the figure inset, is related to the afterpulse probability. We observe a decrease in the afterpulse contribution according to the reduction of the fraction of bright pulses sent by Eve. As Eve attacks more photons, more bright pulses are sent and are consequently measured in the wrong basis by Bob. Despite the fact that these pulses are not capable of triggering an avalanche (due to loss on the PBS) they still generate an electrical current across the diode. The density of trapped charges increases linearly with the current flow through the device when working in linear mode until saturation [37], which leads to an increase in the number of afterpulses. The afterpulse probability values measured for 100%, 50%, 10% and 1% of single-photons attacked by the eavesdropper are 38.54%, 24.13%, 6.96% and 1.77%, respectively, for $2 \times 10^5$ detections in each series. In all cases the eavesdropper intervention can be detected with our monitoring system, as the afterpulse probability is higher than the negligible value for the no-attack case (lower than 0.1%). The relative standard-deviation obtained for five measurement series of $1 \times 10^3$, $5 \times 10^3$ and $4 \times 10^4$ points each, extracted from the data for the full attack condition, was 9.6%, 1.6% and 0.7% respectively. In the case without the imposed deadtime, the afterpulse probabilities were 76.6%, 61.8%, 28.9% and 8.4% respectively, all higher than the reference value of 1.79% without attack.

In addition to the fingerprints left by Eve on the detector parameters, another characteristic of the after-gate attack can be monitored. As Eve only sends her signal after Bob's gate, depending on this delay and on the acquisition board resolution, different time periods between events may be observed in the histogram (in this case, the time bin is given by the sampling rate of the acquisition board). The monitoring of the deviation from the synchronous feature of the gated operation is further explained in section 4, as we employ this principle against the time-shift attack.

Even a small fraction of the attack (1%) remains detectable by the real-time monitoring system, in spite of the 10 μs-deadtime imposed. Since our method is capable of isolating the different sources of detector noise, it is able to pinpoint differences in its expected characteristics, therefore revealing the attack.

*3.2 Faint after-gate attack*

Recently, a slight variation of the after-gate attack has been proposed, called the faint after-gate attack [30]. An interesting feature of the new technique is the usage of weaker pulses by Eve. In the paper, the authors define the term "superlinear threshold detector" for explaining the behaviour of SPAD detection efficiency when the detector is subjected to multiple photons arriving bunched in a small time span or spread along the gate. As the average number of photons per pulse increases, the gate efficiency profile tends to extend, especially at the end. The attack operates similarly to the conventional after-gate, as pulses on two different power levels arrive at the end of the gate of Bob's detectors. For bases agreement between Eve and Bob, the detector exhibits a higher efficiency level due to the pulse's higher optical power; otherwise it exhibits a lower value. Different from the case of the binary response of the conventional after-gate attack, this strategy intrinsically introduces errors on the system [30].

We have experimentally simulated the faint after-gate attack by adjusting the power level of the pulses sent by Eve. The operational point, i.e., the temporal position of the optical pulse inside the detection gate was characterized similarly to our preparation of the conventional after-gate attack. Eve's 1 ns pulses were scanned through Bob's detector gate at different power levels differing by a factor of 3 dB. We have chosen the gate temporal position that has the highest ratio between the high and low normalized efficiency curves at its falling end. The end of the gate is shown in Fig. 4, with the operation point indicated (note that the gate was scanned relative to the optical pulse, so the curves appear mirrored). The efficiencies relative to the gate peak with the optical pulse containing 76 and 38 photons on average at the temporal operation point are 0.0364 and 0.0084, corresponding to a ratio of 4.3 between the higher and the lower efficiencies ($\max\{\eta_{high}/\eta_{low}\}$). The statistical distribution of times between consecutive detection events was once again collected under attack, with Eve intercepting 100% of the photons sent by Alice with detector deadtime removed.

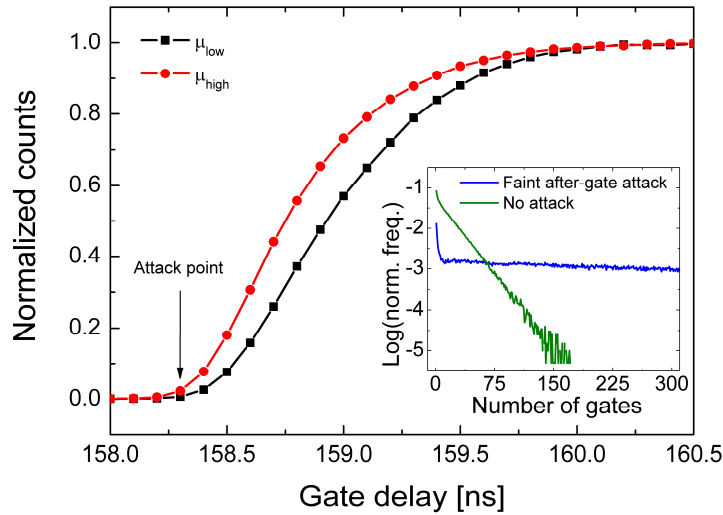

Fig. 4. Gate scan with two different average numbers of photons per pulse for the faint after-gate attack (the end of the gate appears on to left). Inset shows the histograms of times between detections at standard operation and under the attack, with the observable difference in the slopes of the curve due to the fact that when the detector is under attach the detection efficiency is lowered because the attack is performed at the end of the gate.

From the measurements, we found that the detection efficiency observed by Bob is reduced from 15% to 0.4% when under full attack, due to the lower sensitivity at the end of the gate. This reduction is observed as the decrease in the slope of the long tails in the inset of Fig. 4. Eve may compensate this effect by making the link length between Alice and her around 16 dB more transparent, if there is originally such a margin; or Eve may be present during the system synchronization, when the link losses are estimated, and artificially fakes a higher loss, which is removed afterwards. As an example, considering a fiber attenuation of 0.2 dB/km, if Bob's station is 80 km distant from Alice he therefore expects 0.01 photons per pulse, Eve could put the detector right at Alice's output in order to collect 0.4 photons and adjust the faked states optical power to force Bob to detect the expected value of 0.01. We simulate this strategy in the opposite way, by reducing the average number of photons sent directly from Alice to Bob in the standard no-attack case to make a fair comparison of the detector response. Even in this case, Eve leaves fingerprints on Bob's detector, as the afterpulse probability at 500 kHz gate frequency decreased from 5.1% to 2.8% when the detector is under full attack, if compared to standard operation (note that another SPAD, with a higher no-attack reference value for afterpulse probability, was used for this attack). This reduction may be related to the smaller excess voltage of the SPAD bias relative to the breakdown threshold at the end of the gate, resulting in a smaller electronic flux through the device and, as a consequence, a smaller population of traps when compared to full avalanches at the middle of the gate caused by standard operation.

*3.3 Discussion on the performance against general blinding techniques*

Although we have demonstrated that real-time monitoring is effective against the after-gate attack and its faint variation, we would like to discuss the applicability of the method to catch fingerprints left by more general blinding attacks. These blinding attacks may be used on continuously running detectors with CW light [14] through lowering the bias voltage by physically heating the SPAD with the imposed CW light [17], using a combination of CW and pulsed light [20,21] and as recently demonstrated, blinding using weak pulses before the detection gate to explore the detector deadtime [19].

From analyzing the results presented in those papers, we infer that our technique seems to be readily applicable in the case with CW blinding light acting on non-gated detectors [14]. In this case, the large gap required on the CW light reaching the blinded detector, before a count is generated, extends the dead time of the detector artificially. As shown in Fig. 5, which is a measurement of time intervals between consecutive counts generated by a non-gated commercial Si detector, not only can we fully characterize a continuous running detector, we can also monitor the deadtime in real-time. In this case, the time intervals are multiples of the A/D sampling period. As observed in the inset, we measured a deadtime of ~ 40 ns (limited in this case by our A/D resolution of 10 ns), which is typical for actively-quenched non-gated detectors and close to the manufacturer's specification. In [14] the author reports that the gap used is considerably longer than this value (~ 500 ns for an active-quenched model), which should be caught by the monitoring system.

The monitoring scheme may also be used for the attack exploiting the detector deadtime [19]. In that work, Eve sends pulses containing around 16 photons in average before the bit slot defined by the system synchronization. According to the state chosen for each of such pulses, a detector on Bob's station can trigger an avalanche and becomes blinded when Alice's photon arrives, due to the enabled deadtime. One interesting feature of such a strategy is that no interception is performed on the transmitted qubits and Eve does not disturb them, making the attack difficult to catch by QBER analysis or a change in the count rates. Our method could be employed to monitor unexpected time intervals between avalanches, which would indicate an intervention from the eavesdropper (this principle is further explained in the next section, where it is employed against the time-shift attack). The authors proposed that Eve could vary the incidence time of the blinding pulses to make them appear as background noise. Even in this case, our monitoring scheme might be effective by verifying divergences in the overall detection efficiency and dark counts.

For the other attacks mentioned it is not clear if the monitoring system would work since they do not leave obvious traces in the detector's characteristics, such as increased afterpulses, or a different deadtime. They may however leave subtle differences that may be identified, therefore further investigation is needed on these attacks. Nevertheless the monitoring system presents a degree of practical protection for Alice and Bob. Eve needs to replicate the detector's exact characteristics in terms of afterpulses, detection efficiency, dark counts and deadtime when performing these attacks, otherwise Bob will be suspicious that eavesdropping with optical blinding may be occurring.

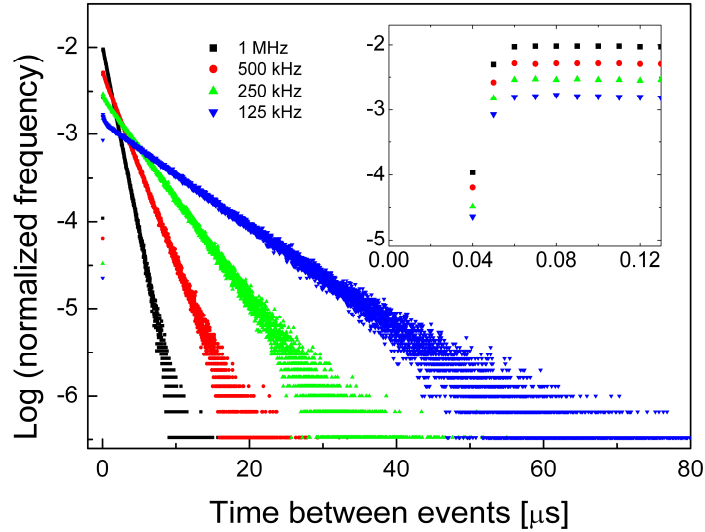

Fig. 5. Histogram of times between consecutive detections for a Silicon SPAD operating in free-running mode for four different single-photon counting rates, as illuminated by an attenuated CW laser source. The inset highlights the detector deadtime.

One possible countermeasure for Bob against these cases, is for him to change a physical parameter of his detector (efficiency or deadtime for instance), from time to time according to a quantum random number generator [38]. Since it is impossible for Eve to know when these parameters have changed, she will not be able to generate a counting distribution that corresponds to the detector with the modified characteristics, and Alice and Bob can be suspicious of an attack. This parameter change can occur fairly often since the monitoring system can reliably characterize the detector with relatively few detections [29]. The main drawbacks of this technique are an increase in QBER due to lower detection efficiency used from time to time and a lower key generation rate when longer dead times are used.

### 4. Monitoring against the time-shift attack

In the time-shift attack [11,12], the system timing is changed in order to exploit the time-dependent mismatch between the efficiency curves of the detectors on the receiver station. Consider a two-detector receiving scheme, as in the case of a polarization-encoded BB84-based setup. Bob has an active element to choose his detection basis and projects each qubit on a polarisation state using a PBS, with each output port connected to a SPAD, $D_A$ and $D_B$. The system clock is synchronized such that the single photons reach $D_A$ at temporal position $\tau_1$ relative to the trigger signal and the device exhibits detection efficiency $\eta_{A1}$. However, Eve randomly routes each transmitted photon through one of two paths of an optical delay line, without interception. For the second delay, corresponding to time $\tau_2$, the efficiency of the detector will be different and smaller ($\eta_{A2}$). Now assume that the efficiency temporal profile for $D_B$ is not matched with $D_A$ and, for time $\tau_2$, the value is high ($\eta_{B2}$), whereas for time $\tau_1$, it is lower ($\eta_{B1}$). This implies in a higher probability that, given a certain basis chosen by Bob, one of the detectors has a greater probability of triggering an avalanche, according to the time-shift imposed by Eve, which will posteriorly provide her with partial information (or full if the mismatch is large enough) about the key after the bases reconciliation.

We performed a proof-of-principle measurement of the times between events recorded on a commercial SPAD when under the time-shift attack. The attenuated signal from a pulsed laser diode (the same one used by Eve in the after-gate attack) is sent to the gated-mode SPAD. The synchronization signal with 100 kHz frequency passes through Eve's FPGA-based, randomly-activated, electrical delay-line and triggers the gate windows (100 ns) on the detector under attack at Bob's station, as shown in Fig. 6.

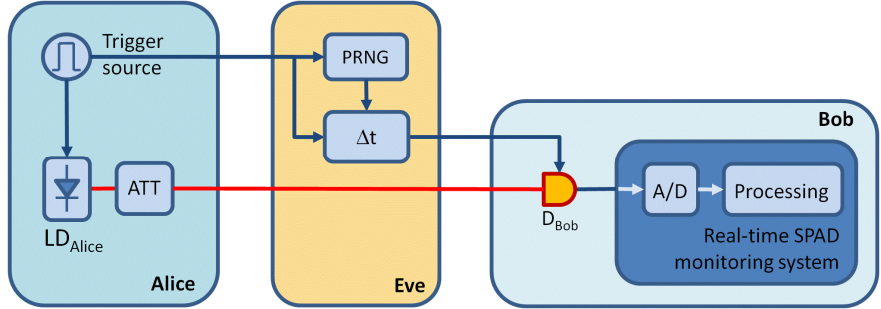

Fig. 6. Setup for the proof-of-principle monitoring against the time-shift attack. Here, the synchronization trigger is delayed instead of the optical pulses.

The electronics occupied by Eve to time-shift the synchronization pulses has a jitter of ~ 20 ns on the electrical pulses feeding Bob´s SPAD. We were forced to employ such a wide window of 100 ns at Bob´s detector to ensure that the attack imposed by Eve on her electronic limitations could be successfully implemented. Note that our A/D acquisition hardware is capable of resolving the time differences imposed by the attack up to the order of 10 ns. Many modern QKD setups already employ much faster electronics for high-speed detection and basis choice [39] and thus it should be possible to catch attacks imposed on typical QKD gate windows with the same hardware already present, or at most with an upgrade in the detection electronics. For half of the total number of pulses, a relative delay of 60 ns is imposed to the signal, simulating the attack. The normalized temporal profile of the SPAD gate, measured with the bypassed, fully delayed and randomly delayed version of the trigger signal is shown in Fig. 7. The operational point for the attack was set at the position of the traced line.

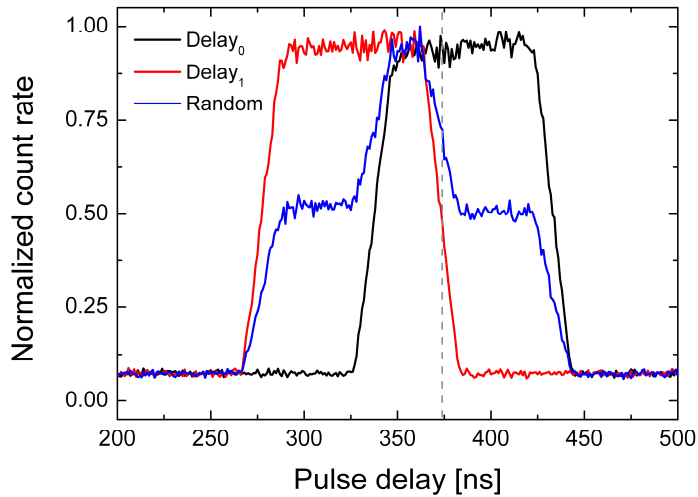

Fig. 7. Temporal scan of the detection gate with the delayed optical pulse. The operational point is identified by the traced line.

This point corresponds to a ratio of 50% between the lower and higher efficiency values explored by Eve (red curve). Here we assume that the other detector at Bob's station (not used here) has a complementary behaviour, for example, similar to the black curve in Fig. 7. This results in a mutual information value between Eve and Bob of 0.082 [11], which tends to unity if the efficiency ratio approaches zero.

The SPAD output signal feeds a 100 MSamples/s A/D board with an internal time base, in order to be able to resolve the different time peaks induced by the attack, and the time elapsed between consecutive detections is recorded (as multiples of the sampling period, 10 ns). The normalized histogram of times is presented in Fig. 8 for the case when the detector is under attack and for a total bypass of the random delay (or when all pulses are delayed), with no deadtime.

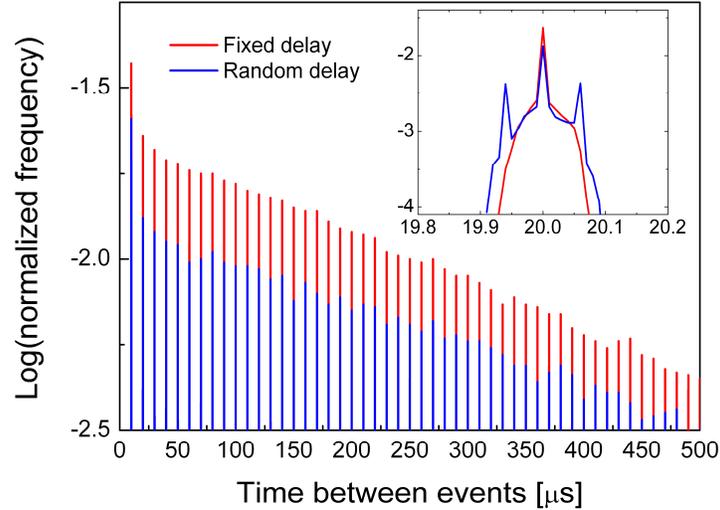

Fig. 8. Histogram of times between consecutive events when under the time-shift attack (blue lines) and under normal operation. Each time bin is defined by the A/D resolution. Inset zooms in the second bin.

From Fig. 8, we see the discretized periodic behavior of the detector, due to the gated-mode operation, when the analysis is based on the sampling period. According to the delay values set by Eve, three results are possible for the time between two consecutive events: if any two consecutive detection events are (or not) delayed, the histogram peak corresponds to a multiple of the gate period. If the first event is delayed and the second one is not, the time between them is shortened by the relative time delay. In the opposite situation, the time delay is enlarged by the same time span. This results in a three-peak-per-bin histogram.

The mismatch between the efficiency values at both temporal positions inside the detection gate is the core of the attack strategy. As a consequence, the central peak seen in the figure inset is higher, according to the ratio of efficiencies. In addition, the efficiency mismatch is experienced by Bob as a drop in the net efficiency of his detectors, which can also be visualised as an increase in photon loss. This effect appears on the histogram of times as a slight reduction in the slope of the long curve tail (in log scale). The net efficiency extracted from the histograms presents a drop from 15.0 (for the case of no attack) to 11.8% when under the time-shift attack. From Eve's viewpoint, differently from the after-gate attack, as the photons are not intercepted and resent, it tends to be more difficult to compensate for this loss. If Eve uses currently available technology, the decreased net efficiency due to the attack, observed by Bob as a reduction in average number of photons, cannot be compensated, unless the optical channel is replaced by a more transparent one. Even in this case, when the attack is not directly identified by the histogram statistics, our method would be capable of resolving the time-shifts, as shown in the Fig. 8 inset, as long as the resolution of the analyzing electronics is sufficient. Eve may of course attempt to use a shorter delay shift in order to avoid being caught, but there is a minimum bound to this, as it is limited by Bob's detectors jitter. In any case, this result is sufficient to demonstrate that our monitoring method is capable of identifying non-expected timing changes, without tampering on the detector, but only analysing its electric output.

## 7. Conclusion

We demonstrated that on-line monitoring of the SPADs on a QKD system can be used to detect the after-gate and the time-shift attacks without tampering internally with the single-photon detector. This tool is a countermeasure that Alice and Bob may use, with the advantage that only changes in the detection electronics are needed, thus no increased optical loss at Bob's station and no extensive hardware modifications required, further motivating its application in real-world scenarios.

The after-gate attack was demonstrated in a simplified QKD polarization-encoded setup and the detector was monitored for different ratios of intercepted per transmitted photons. The histogram of times between consecutive detection events with the SPAD under attack was measured and compared with the standard operational mode. We verified a measurable increase in the afterpulse probability even for a very small (1%) fraction of photons being eavesdropped. The analysis of the characterization results for different statistical sample lengths demonstrated the feasibility of our method detecting the after-gate attack. The faint after-gate attack was similarly detected with the same analysis as in the standard after-gate case when the detector deadtime is removed.

We have also discussed that the monitoring system may offer protection against more general blinding attacks. In this situation the attack also becomes harder for a practical Eve, as she needs to replicate the detector's counting statistics. A countermeasure becomes possible in such cases if a physical parameter change of Bob's detectors is performed at random times, such as its efficiency.

A proof-of-principle for our method against the time-shift attack was conducted. For systems working under synchronous gated-mode, the real-time monitoring is capable of providing temporal information about the relative delay imposed by Eve on the photons sent by Alice, if the acquisition sampling rate is adequate.

Quantum hacking is an emerging area and will become more sophisticated as new techniques are developed exploiting the physical properties of the different components of a QKD system. The ability to constantly monitor single-photon detectors in real-time and under operation during a quantum key exchange provides Alice and Bob with an extra tool to detect different types of hacking strategies.

## Acknowledgments

This work is supported by Brazilian agencies CAPES, CNPq and FAPERJ. In addition G. B. Xavier acknowledges support from grants Milenio P10-030-F, CONICYT PFB08-024 and FONDECYT no. 11110115. Authors thank G. Amaral for help with the FPGA.